\documentclass[12pt]{article}

\usepackage{rotating}
\usepackage{graphicx}
\usepackage{setspace}
\usepackage{calc}
\usepackage{latexsym,amssymb,amsmath}

\newcommand{\xavr}[1]{<\!\!#1\!\!>}

\newcommand{\bequ}{\begin{equation}}
\newcommand{\eequ}[1]{\label{#1}\end{equation}}
\newcommand\eq[1] {(\ref{#1})}
\newcommand{\bfm}[1]{\boldsymbol{#1}}

\begin{document}
\title{Global communications in multiprocessor simulations of flames}
\author{V. Karlin\thanks{University of Central Lancashire,Preston PR1 2HE, United Kingdom}
\thanks{VKarlin@uclan.ac.uk}
\thanks{http://www.uclan.ac.uk/facs/destech/builtenv/vkarlin/}}
\date{}

\maketitle
\thispagestyle{empty}

\begin{abstract}
In this paper we investigate performance of global communications 
in a particular parallel code. The code simulates dynamics of 
expansion of premixed spherical flames using an asymptotic model 
of Sivashinsky type and a spectral numerical algorithm. As a 
result, the code heavily relies on global all-to-all 
interprocessor communications implementing transposition of the 
distributed data array in which numerical 
solution to the problem is stored. This global data 
interdependence makes interprocessor connectivity of the 
HPC system as important as the floating-point power of the 
processors of which the system is built. Our experiments 
show that efficient numerical simulation of this particular 
model, with global data interdependence, on modern HPC systems 
is possible. Prospects of performance of more sophisticated 
models of flame dynamics are analysed as well.
\end{abstract}


\section{Introduction} 

The problem under consideration stems from the studies of 
dynamics of premixed 
flames affected by the hydrodynamic flame instability. These 
flames are common in science and engineering applications, in 
particular in working out industrial safety requirements in 
petrochemical plants, in design of internal combustion engines, 
and in astrophysics. Expanding spherical flames are one of the 
most important types of premixed flames and are 
a popular subject of fundamental and applied studies. In particular, a 
possibility of a self-induced transition from deflagration 
to detonation is the issue of a 
great importance both in practice and theory.

One of the most distinctive features of these flames 
is that their expansion rate 
does not remain constant, as one would expect, but it grows 
in proportion to the square root of time 
\cite{Gostintsev-Istratov-Shulenin88}. Theoretical 
evaluations indicate that the expansion rate should 
depend on physical dimension of the problem and numerical 
simulations confirmed this for planar circular flames, 
see e.g. \cite{Karlin-Sivashinsky05a} and references therein. 
Calculations of spherical flames carried out so far, using 
an asymptotic Sivashinsky type model, remain inconclusive, 
because it was impossible to cover long enough time intervals 
to allow flames to stabilize to the asymptotic power-law 
regime \cite{Karlin-Sivashinsky05b}. The direct three-dimensional 
Navier-Stokes simulations do not look feasible at present at all, 
because long time intervals mean large flame radii, i.e. large 
Reynolds numbers. 

Mathematical model used in our simulations needs 
intensive all-to-all data exchanges and, as a result, the 
scalability of our MPI code is expected to degrade as the 
number of processors brought into play grows. Extrapolation 
of the code performance has indicated that the computer 
resources needed to carry out the calculations until 
stabilization of flames to the asymptotic power-law regime 
may require up to a few processor-years. On the other hand, 
reaching this asymptotic power-law regime of expansion of 
spherical flames numerically is crucial in order to justify 
studies of pseudoresonant interaction of flame wrinkles with 
the upstream velocity perturbations \cite{Karlin05} as a 
possible mechanism of triggering transition of deflagration 
to detonation. 

Modern massively parallel computers differ from each other 
by design substantially. Accordingly, performances of their 
inter-processor communications differ as well. Hence, 
some HPC systems and algorithms of global interprocessor 
communications might be more beneficial for one particular 
type of a parallel code than for another. In this paper we 
describe results of porting our Fortran-90/MPI code on SGI 
Altix 3700, NEC SX-8, and Cray XT4 and analyze its comparative 
performance for a variety of programming implementations of 
global communications.

In the following Section \ref{Model} we outline the 
governing equation and numerical algorithm to solve it. 
Typical results of numerical experiments with expanding 
flames are illustrated in this section too.
In Section \ref{Code} we describe the structure of our 
computational code. Results of the benchmark runs are 
presented in Section \ref{Three}.

\section{Mathematical Model and Numerical Algorithm}\label{Model}

Here we study computational performance of a parallel code 
for numerical simulation of dynamics of three-dimensional 
flames using the asymptotic model suggested in 
\cite{Karlin-Sivashinsky05b}, where a review of other 
available models can be found too. 

Let us consider an expanding flame front and assume that 
its surface is close enough to a sphere and that every point 
on the flame surface is uniquely defined by its distance 
$r=r(\theta,\phi,t)$ from the origin for $0\le\theta\le\pi$, 
$0\le\phi\le 2\pi$, and $t>0$. It is convenient to represent 
such a flame as a perturbation $\Phi(\theta,\phi,t)$ of a 
spherical surface of a reference radius $r_{0}(t)$, i.e.
$r(\theta,\phi,t)=r_{0}(t)+\Phi(\theta,\phi,t)$. Denoting 
Fourier components of $\Phi(\theta,\phi,t)$ as 
$\widetilde{\Phi}_{\bfm{k}}$, the Fourier image of the 
equation governing evolution of the three-dimensional 
segments of such flames can be written as
\[
\frac{d\widetilde{\Phi}_{\bfm{k}}}{dt}
=\left\{-\frac{\theta_{\pi}^{2}}{[r_{0}(t)]^{2}}|\bfm{k}|^{2}
+\frac{\gamma\theta_{\pi}}{2r_{0}(t)}|\bfm{k}|\right\}\widetilde{\Phi}_{\bfm{k}}
\]
\bequ
-\frac{\theta_{\pi}^{2}}{2[r_{0}(t)]^{2}}\sum\limits_{\bfm{l}\in\mathbb{Z}^{2}}
\bfm{l}\cdot(\bfm{k}-\bfm{l})\widetilde{\Phi}_{\bfm{l}}\widetilde{\Phi}_{\bfm{k}-\bfm{l}}
+\widetilde{f}_{\bfm{k}}(t),\qquad t>t_{0}>0,
\eequ{SivCurv1c}
where $\bfm{k}=(k_{1},k_{2})$, $\bfm{l}=(l_{1},l_{2})$ are pairs 
of integers such that $-\infty<k_{i},l_{i}<\infty$, $i=1,2$, 
which can be denoted also as $\bfm{k},\bfm{l}\in\mathbb{Z}^{2}$, 
$\widetilde{f}_{k}(t)$ are the Fourier components of the properly 
scaled upstream perturbations of the unburnt gas velocity field 
$f(\phi,t)$, $\gamma=1-\rho_{b}/\rho_{u}$ is the contrast of 
densities $\rho_{b},\rho_{u}$ of the burnt and unburnt gases 
respectively, and initial values of
$\widetilde{\Phi}_{k}(t_{0})=\widetilde{\Phi}_{k}^{(0)}$ are 
given. By construction, \eq{SivCurv1c} governs dynamics of 
equatorial flame segments 
$-\pi/(2\theta_{\pi})\le\theta\le \pi/(2\theta_{\pi})$, 
$0\le\phi\le 2\pi/\phi_{\pi}$, where $\theta_{\pi}$, 
$\phi_{\pi}$ are integers, see \cite{Karlin-Sivashinsky05b}. 

For the unity Lewis number the length and time scales used in 
\eq{SivCurv1c} are the thermal flame front width 
$\delta_{th}=D_{th}/u_{b}$ and $\gamma^{-2}\delta_{th}/u_{b}$ 
respectively, where $D_{th}$ is the thermal diffusivity of the 
system and $u_{b}$ is the planar flame speed relative to the 
burnt gases. Choice of $r_{0}(t)$ in the model just introduced 
may be based on a variety of principles, here we use 
$r_{0}(t)=t$. 

System \eq{SivCurv1c} is solved numerically by neglecting 
the harmonics of orders higher than a finite integer number 
$K>0$. Then, the nonlinearity can be represented as a circular 
convolution and evaluated effectively with the FFT. The 
structure of the two-dimensional 
FFT, as the tensor product of the one-dimensional ones, allows 
for efficient parallelization of the computational algorithm. 
Further details of the numerical integration algorithm can be 
found in \cite{Karlin-Sivashinsky05a,Karlin-Sivashinsky05b}.

\begin{figure}[ht]
\centerline{\includegraphics[height=72mm,width=144mm]{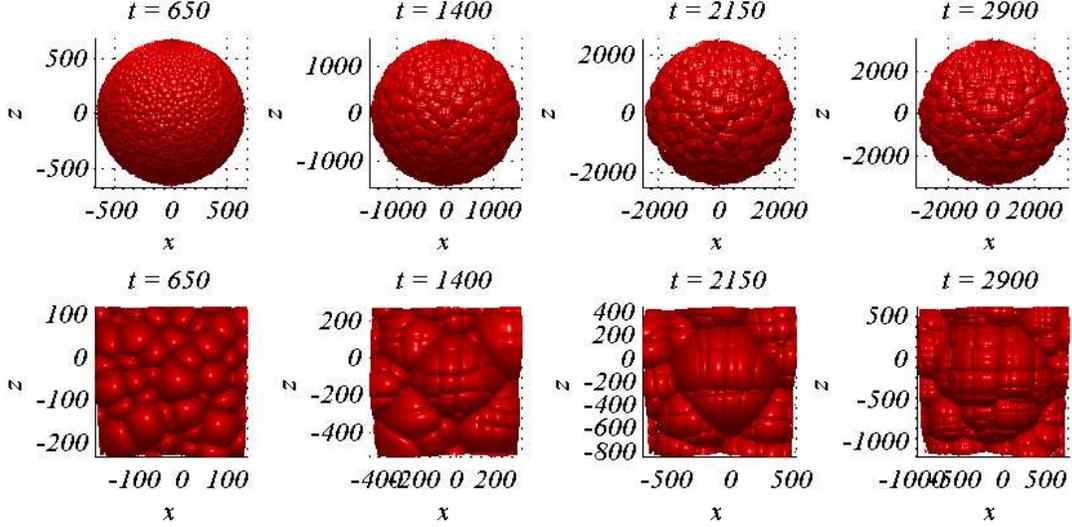}}
\caption{Evolution of half of a spherical flame (top row) and 
of its centrally located region (bottom row). The view in 
the latter sequence is limited by a cone with fixed apex angle 
in the origin. Here $\gamma=0.8$ and $f(\phi,t)\equiv 0$.}\label{ThreeGen}
\end{figure}

In spite of drastic simplifications made during evaluation 
of the three-dimensional asymptotic model \eq{SivCurv1c}, its 
numerical 
simulations are still extremely computationally intensive and 
computational resources available to us allowed to consider 
only a few cases so far. In the case presented below, the  
initial perturbations of the spherical flame surface were 
random $\Phi(\theta,\phi,t_{0})=\epsilon\eta(\theta,\phi)$. 
Here $|\eta(\theta,\phi)|\le 1$ is a number chosen randomly 
for every required combination of $\theta$ and $\phi$, 
$\epsilon=10^{-3}$ and $t_{0}=200$. The explicit forcing was 
not applied in these examples, though the round-off errors 
were playing its role implicitly. Figure \ref{ThreeGen} 
shows evolution of half of a spherical flame in the top and 
of its central region in the bottom. Cells in the flame 
regions near the poles appear severely deformed, because of 
a simplified treatment of the spherical geometry in 
\eq{SivCurv1c}, and should be disregarded. 

\begin{figure}[ht]
\centerline{\includegraphics[height=42mm,width=55mm]{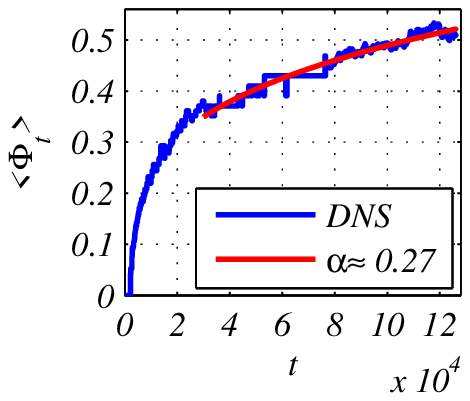}
$\left.\hspace{10mm}\right.$
\includegraphics[height=42mm,width=95mm]{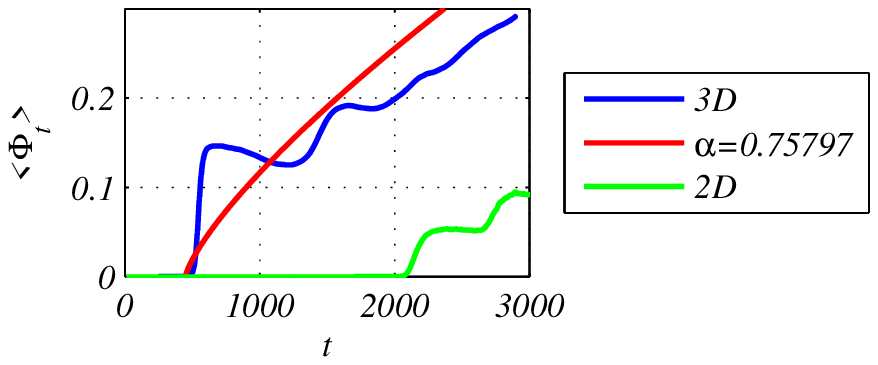}}
\caption{Averaged flame front velocities and their power law 
approximations $(t-t_{*})^{\alpha}$ for circular (left) and 
spherical (right) flames versus time $t$ (right). Here 
$\gamma=0.8$ and $f(\phi,t)\equiv 0$.}\label{SphereTwoD}
\end{figure}

Averaged velocities of the spherical flames $\xavr{\Phi_{t}}$ 
and their power law approximations 
$(t-t_{*})^{\alpha}$ are depicted in Fig. \ref{SphereTwoD} 
as well. As time goes by, the expansion rate of the 
two-dimensional circular flame stabilizes to 
$\xavr{\Phi_{t}}\propto(t-t_{*})^{1/4}$, i.e. to the 
power law predicted by the fractal analysis. The 
three-dimensional flames accelerate much faster than the 
two-dimensional ones. The power law approximation of the 
expansion rate on the time interval considered so far 
gives $\alpha\approx 0.76$, indicating good chances that 
the expansion rate will eventually stabilize to 
$\xavr{\Phi_{t}}\propto(t-t_{*})^{1/2}$ after some transitional 
period as demonstrated by the experiment and predicted by 
fractal analysis. 

Further analysis of results of numerical simulations can be 
found in \cite{Karlin-Sivashinsky05a,Karlin-Sivashinsky05b}.

\section{The structure of the computational code}\label{Code}

The main computational array of the problem is 
$\widetilde{\Phi}_{k_{1},k_{2}}$, where $-K\le k_{1},k_{2}\le K$ 
and $K$ is a large positive integer. In order to tackle the 
nonlinear term in \eq{SivCurv1c} as a circular convolution, the 
array $\widetilde{\Phi}_{k_{1},k_{2}}$ is augmented to 
$\Psi_{k_{1},k_{2}}$, where 
$-2K\le k_{1},k_{2}\le 2K+k_{0}$, see e.g. \cite{VanLoan}. The 
size of the augmented array is $(4K+k_{0}+1)\times(4K+k_{0}+1)$ 
and integer $k_{0}$ is selected as minimal as possible to make 
value of $K_{\Psi}=4K+k_{0}+1$ efficient for the FFT. 

Initially, the augmented array 
$\Psi_{k_{1},k_{2}}$ is distributed 
between $N_{p}$ processors column-wise, so that its first 
$K_{p}=K_{\Psi}/N_{p}$ columns are 
stored in the first processor, the next $K_{p}$ columns are 
in the second processor and so on. At this stage, the 
one-dimensional FFT can be effectively applied along the first 
index of the arrays $k_{1}\Psi_{k_{1},k_{2}}$ and 
$k_{2}\Psi_{k_{1},k_{2}}$, i.e. 
column-wise. In the next stage, the resulting global arrays 
should be transposed in order to accomplish the two-dimensional 
FFT by applying the one-dimensional FFT along the second index, 
i.e. raw-wise.

\begin{figure}[ht]
\centerline{\includegraphics[height=160mm,width=50mm]{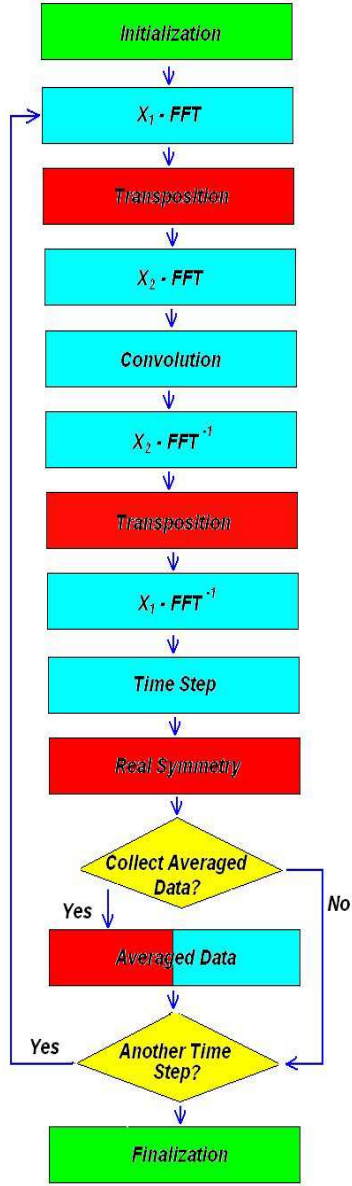}}
\caption{The block-scheme of the code. Red blocks are communication 
and cyan ones are floating-point operations intensive.}\label{Block}
\end{figure}

Upon completion of 
the two-dimensional FFT, the convolution is calculated as the 
element-by-element product of two FFT'ed arrays and the just 
describer set of operations is repeated in the backward 
direction. First, the one-dimensional inverse FFT is applied 
along the second index of the Fourier image of the convolution. 
Then, the global array is transposed back to its column-wise 
distribution and the one-dimensional inverse FFT is applied 
along the first index.

The overall structure of the code is illustrated in the block 
diagram in Fig. \ref{Block}. Interprocessor communications 
intensive parts of the code correspond to rectangles with 
smoothed vertices and floating-point arithmetics intensive 
modules are denoted by standard rectangles. Computational 
resource consumption in blocks not shown in the diagram is 
not essential. The code was written in Fortran-90 and MPI. 
Vendor recommended FFT routines were used on all tested HPC 
systems.

The problem of transposition of data arrays distributed among 
a number of processors is quite common in scientific and 
engineering applications of HPC. A number of strategies and 
algorithms were suggested for various computer architectures 
and data structures, see e.g. 
\cite{Swarztrauber98,AlNa'mneh-Pan_Yoo06}. Here, we are 
considering the simplest 
column- and raw-wise distributions of two-dimensional arrays only 
and study their transpositions by straightforward MPI tools on 
a few popular HPC systems.

The straightforward implementation of transposition of 
distributed data arrays can be carried out using the 
\verb#MPI_ALLTOALL# routine. Before the call to the 
\verb#MPI_ALLTOALL#, local arrays should be reshaped, 
because the \verb#MPI_ALLTOALL# routine requires the data 
destined for a particular processor to be stored in RAM 
continuously. After the call, the newly assembled local 
arrays should be reshaped back, now in the raw-wise manner in 
terms of the reference distribution of the global array, 
in order to apply the one-dimensional FFT along the second 
index. The entire procedure trebles the required memory, 
but RAM is usually less critical in the problem in question 
than the interprocessor communications efficiency. This 
transposition procedure is illustrated in Fig. \ref{Transp} 
and the code itself is given in the Appendix. 

\begin{figure}[ht]
\centerline{\includegraphics[width=150mm,height=100mm]{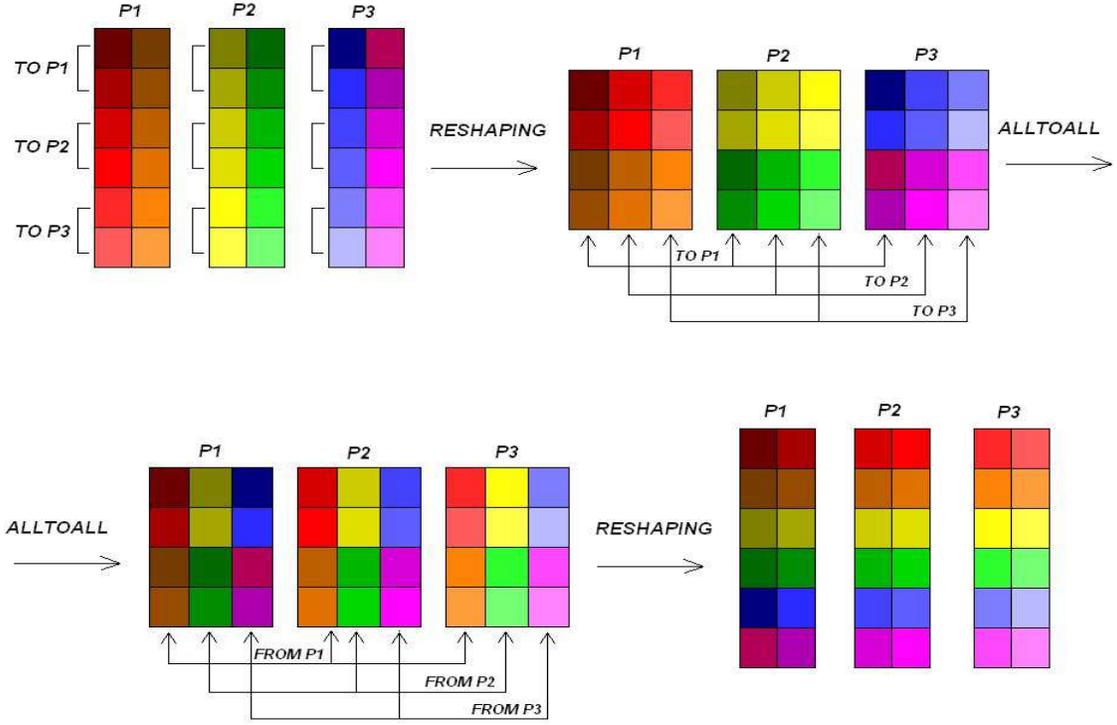}}
\caption{The global data array transposition procedure. 
The illustration features a $6\times 6$ array distributed 
over 3 processors $P_{1}$, $P_{2}$, and $P_{3}$.}\label{Transp}
\end{figure}

The \verb#MPI_GATHERV# routine can be called by every processor in 
order to fetch missing $(N_{p} - 1)K_{p}^{2}$ elements of 
$\Psi_{k_{1},k_{2}}$ and form its raw-wise representation as an 
alternative to the use of the \verb#MPI_ALLTOALL# routine in the
implementation of the global array transposition. This approach 
avoids reshaping of local arrays and is RAM economical. However, 
in our experiments all tested HPC systems performed faster with 
the transpositions based on the \verb#MPI_ALLTOALL# routine. 
As RAM is usually less critical to the problem in question 
than the interprocessor communications efficiency, the latter 
approach appears inferior to the former one on HPC's with 
sufficient memory resources. 
Accordingly, in what follows we present results for the algorithm 
based on the \verb#MPI_ALLTOALL# routine only.

The transposition of the global array requires in our case 
cross-processor transmission of 
$N_{comm}\propto K_{\Psi}^{2}(1-N_{p}^{-1})$ 
elements of $\Psi_{k_{1},k_{2}}$ and is the most communication 
resource-demanding segment of the code. Combined use of the FFT 
routine requires $N_{fp}=\mathcal{O}(K_{\Psi}^{2}\log_{2}K_{\Psi})$
floating-point operations per time step making it the most 
critical arithmetic resource consumer. Thus, the ratio of numbers 
of floating point to communication operations is  
$N_{fp}/N_{comm}=\mathcal{O}(\log_{2}K_{\Psi}/(1-N_{p}^{-1}))$, 
i.e. practically $N_{fp}/N_{comm}\approx 1$ for $N_{p}\gg 1$. In 
other words, speed of communications is as important as the speed 
of arithmetic operations. This is in startling contrast with 
explicit approximations of local PDE's like Navier-Stokes or 
Maxwell systems. In the latter case explicit approximations of 
local derivatives would require communications just between 
neighbouring processors involving grid nodes from close 
proximities of their common boundaries only. 
Hence, the numbers of communication and arithmetic operations 
will be $N_{comm}=\mathcal{O}(K_{\Psi}N_{p})$ and $N_{fp}=\mathcal{O}(K_{\Psi}^{2})$, 
resulting in $N_{fp}/N_{comm}=\mathcal{O}(K_{\Psi}/N_{p})$ and meaning 
that speed of interprocessor communication is not that important 
as speed of arithmetic operations for $N_{p}\ll K_{\Psi}$.

\begin{figure}[ht]
\centerline{\includegraphics[width=90mm,height=60mm]{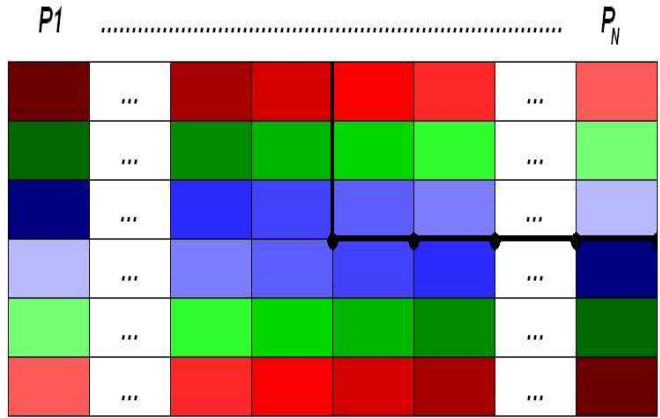}}
\caption{The global data array symmetrization procedure. Data 
from the upper-right corner are spread to the rest of the array 
according to the real symmetry relation 
$\Psi_{-k_{1},-k_{2}}=\overline{\Psi_{k_{1},k_{2}}}$.}\label{Symmetry}
\end{figure}

Another interprocessor communications intensive routine of our 
code is the symmetrization of the global solution array in 
order to ensure it remains real valued in physical space. 
Programmatically this is carried out by mapping the upper half 
of the global array onto the lower one in accordance with the 
symmetry relationship
$\Psi_{-k_{1},-k_{2}}=\overline{\Psi_{k_{1},k_{2}}}$, see 
Fig. \ref{Symmetry}. 
The routine transmits roughly half of the array 
$\Psi_{k_{1},k_{2}}$ and is called once per time step, which 
makes it less resource demanding than the transposition one. 
It is built of the non-blocking point-to-point communications 
subroutines \verb#MPI_ISEND# and \verb#MPI_IRECV#.
Similarly, the numerical integration to advance the 
solution in time needs some floating-point arithmetics, 
but its resource requirements are inferior to the FFT's.

The routine, which calculates a variety of averages of 
obtained numerical solution does both interprocessor 
communications, by using \verb#MPI_ALLREDUCE#, and 
floating-point arithmetics. Again, its resource demands are 
inferior to the crucial blocks identified above. In 
addition, calls to this routine are not compulsory at every 
time step.

\section{Comparative performance of the code}\label{Three}

The Fortran-90/MPI code underwent some basic optimization during 
porting on a particular HPC system. Such 
basically optimized clones of the code were run on a set of HPC's 
with varying numbers of processors and solution array sizes. A few 
runs of the original non-optimized code have been carried out too in 
order to illustrate the value of adjustments to particular HPC 
architecture. The NEC SX-8 required the largest 
optimization effort, though the pay off was the most ample too.

\begin{figure}[ht]
\centerline{\includegraphics[width=70mm,height=56mm]{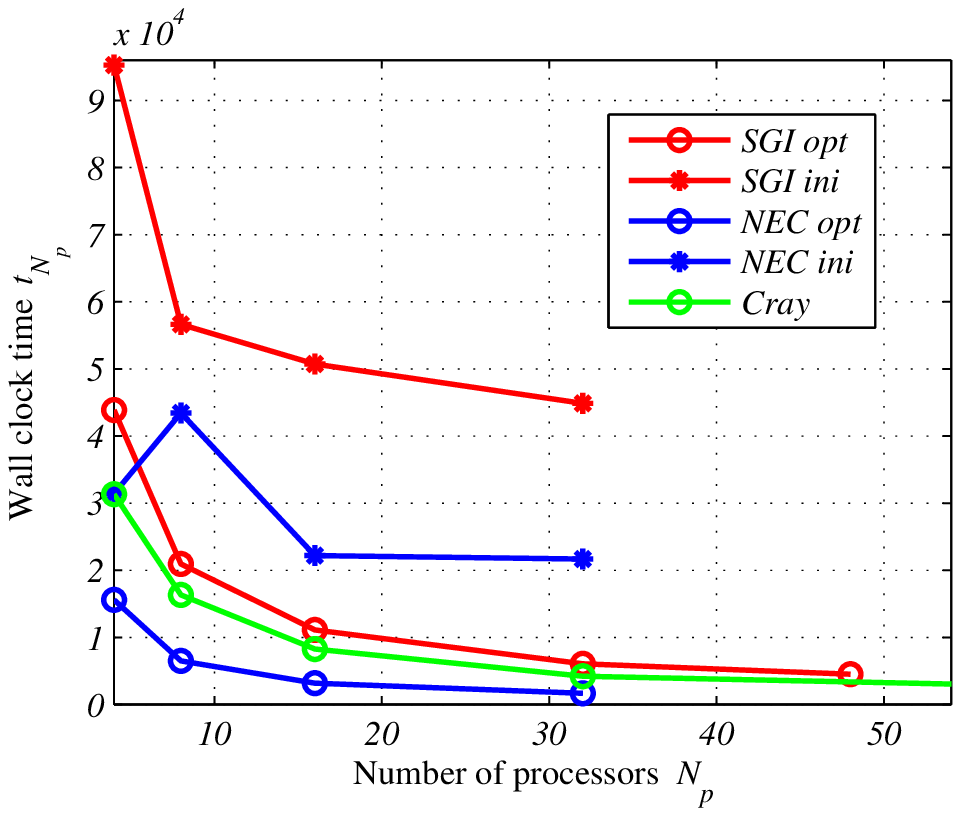}
$\left.\hspace{10mm}\right.$
\includegraphics[width=70mm,height=56mm]{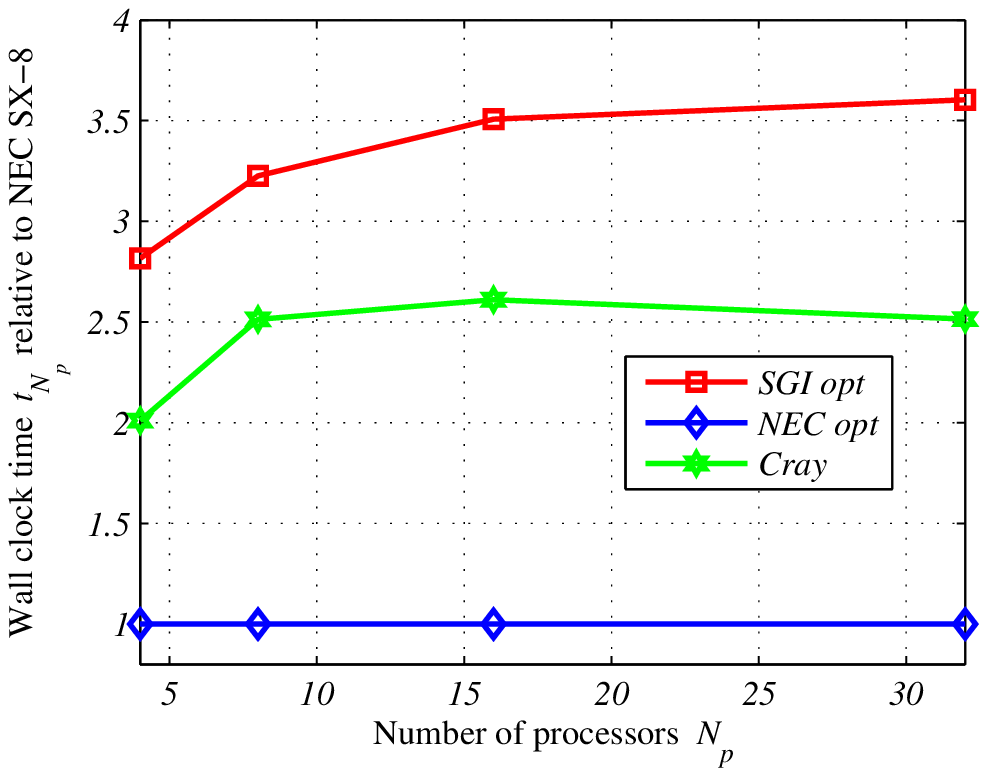}}
\caption{Comparison of the wall clock time $t_{N_{p}}$ 
needed to complete the task (left) and their ratios 
to the wall clock time for NEC SX-8 (right) versus number of 
processors $N_{p}$. Tag \textit{ini} corresponds to an initial 
non-optimized code, its absence - to the optimized ones.}\label{HPC1}
\end{figure}

Graphs in Fig. \ref{HPC1} show that the code in question 
runs on the SX-8 on average about three times faster than on Altix 
and about twice faster than on XT4. Further, they  
reveal that the optimized code is perfectly scalable in spite 
of its global all-to-all communications. Tests were carried out 
with $K_{\Psi}$ ranging from 2000 to 32000 and the graphs 
correspond to $K_{\Psi}\approx 8000$. Variations of $K_{\Psi}$ 
in the given limits does not affect behaviour of graphs in 
relation to each other significantly.

\begin{figure}[ht]
\centerline{\includegraphics[width=70mm,height=56mm]{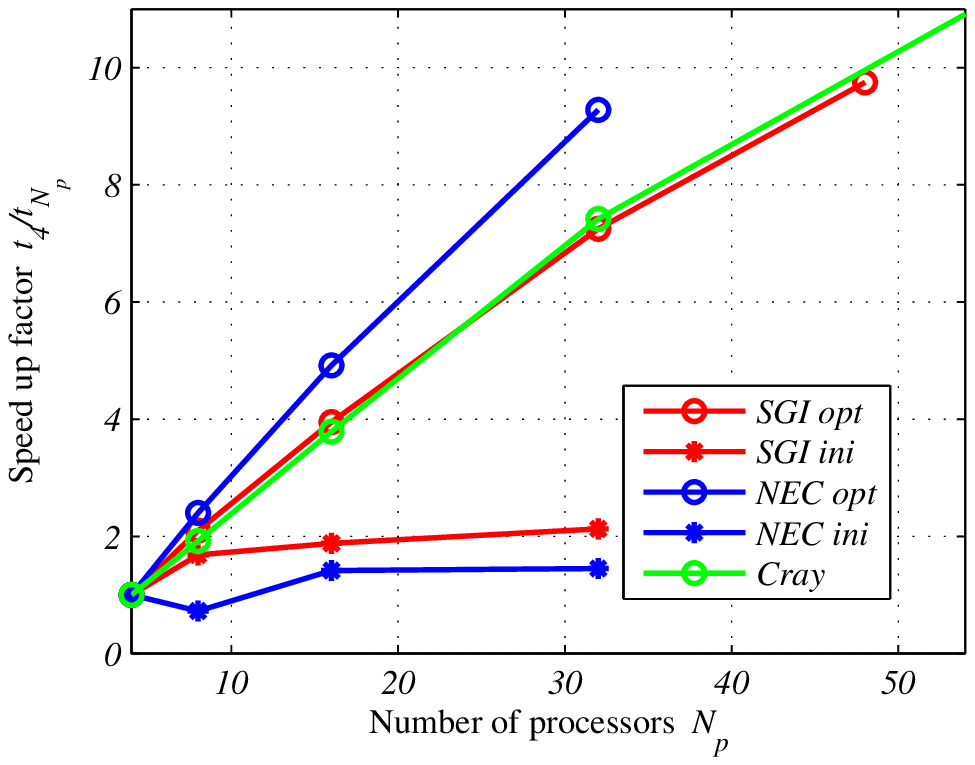}
$\left.\hspace{10mm}\right.$
\includegraphics[width=70mm,height=56mm]{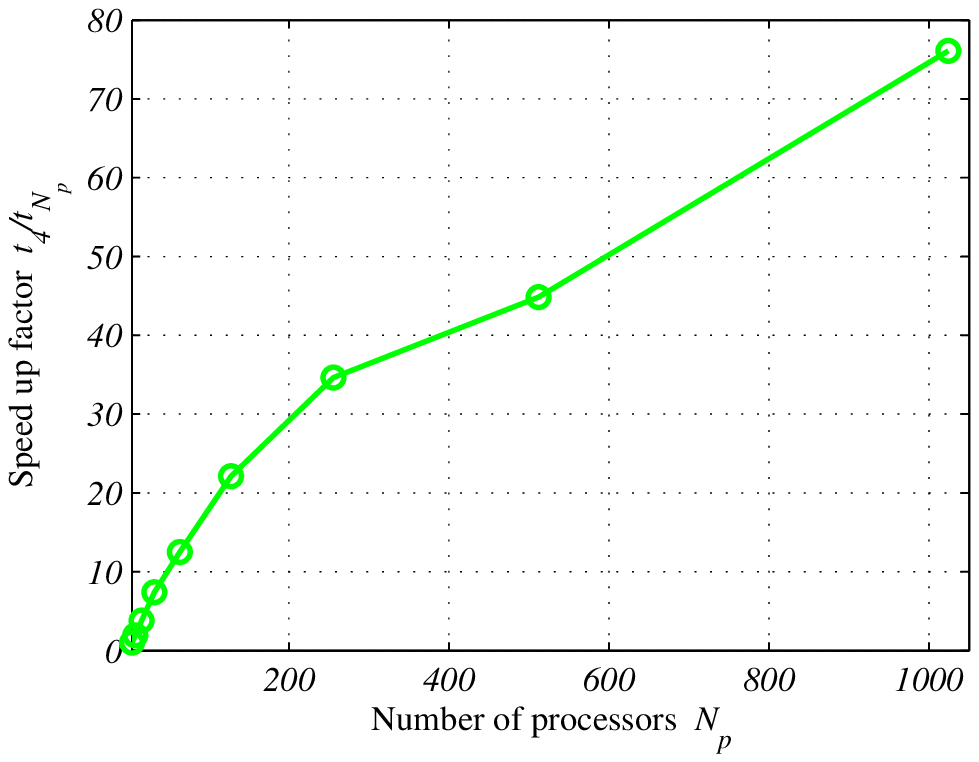}}
\caption{Comparison of the speed up factor $t_{N_{p}=4}/t_{N_{p}}$ 
versus number of processors $N_{p}$. Tag \textit{ini} corresponds 
to an initial non-optimized code, its absence - to the optimized 
ones.}\label{HPC2}
\end{figure}

Our calculations on Altix were practically limited by 48 
processors and by 32 processors on SX-8. However, it was 
possible to use up to 1024 processors of XT4. Corresponding 
graph is depicted in Fig. \ref{HPC2}. It reveals, that scalability 
of the code starts to degrade from $N_{p}\approx 128$. However, 
even for $N_{p}=1024$ the code manages to utilize at least 
40\% of the added computing power.

Profiling shows that further improvement of the code performance 
on all tested systems is impeded by the \verb#MPI_ALLTOALL# 
routine and to a lesser degree by the FFT procedures.
Results of comparison of performances of our particular code on 
a set of HPC's should not be interpreted
as the overall superiority of one of those HPC's over others.

\begin{figure}[ht]
\centerline{\includegraphics[width=70mm,height=70mm]{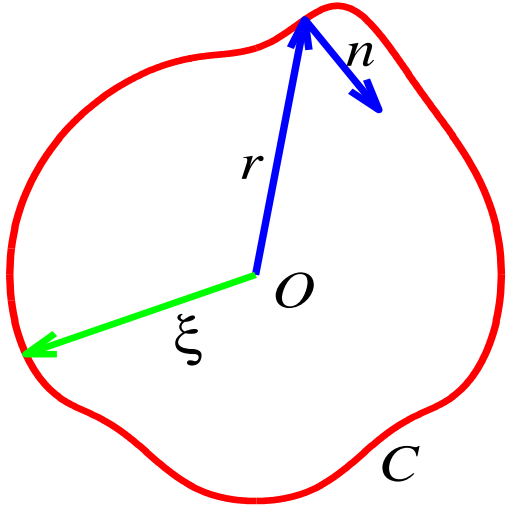}}
\caption{Geometry of the planar flame front $C$ propagating outwards. 
$\bfm n$ is the inner unit normal.}\label{Frankel}
\end{figure}

Being computationally efficient and physically plausible, 
the Sivashinsky type models are not free from drawbacks. For 
example, they neglect certain types of low order spherical 
harmonics and are not valid for the entire expanding flame 
front, see e.g. \cite{Karlin-Sivashinsky05b}. 
Global geometrically correct asymptotic models were 
suggested as well, see e.g. \cite{Frankel90}. In planar geometry
the governing equation of such a model is as follows
\bequ
\frac{d\bfm r}{dt}=\left[
-1+\epsilon\kappa(\bfm r,t)+\frac{\gamma}{2}\left(1+\frac{1}{\pi}
\int\limits_{C}\frac{(\bfm r-\bfm\xi)\cdot\bfm n(\bfm r,t)}
{|\bfm r-\bfm\xi|^{2}}dl_{\bfm\xi}
\right)\right]\bfm n(\bfm r,t).
\eequ{FraCurv1c}
Here $\epsilon$ and $\gamma$ are physical constants, $\kappa$ 
is the curvature and $dl_{\bfm\xi}$ is the increment of length of 
contour $C$. Other notations are explained in Fig. \ref{Frankel}.

Unlike the Sivashinsky models, the right-hand side of the latter 
ones cannot be represented 
as a tensor product of one-dimensional operators precluding 
use of efficient column/row-wise data structure and causing further 
problems in efficient parallelization. However, its singular 
integral in the right-hand side can be treated with the fast 
multipole techniques \cite{Greengard-Rokhlin87} reducing the number 
of required floating point operations per time step to roughly 
$N_{fp}=\mathcal{O}(K^{2})$. The number of required communications 
per time step will be in 
the range $\mathcal{O}(K^{2})<N_{comm}<\mathcal{O}(N_{p}K^{2})$, 
meaning that the efficiency of computations with this model will 
be defined by the quality of programming and acceptable compromise 
with accuracy. 

On the other hand, explicit economical discretization of the 3-D 
Navier-Stokes system would lead to $N_{fp}=\mathcal{O}(K^{3})$, and 
approximately the same number of communications. It should be noticed 
however, that approximation of the Navier-Stokes system in the 
radial direction would need much larger number of grid nodes 
than in the azimuthal ones, $K$, in order to resolve the structure of 
the flame front. Furthermore, the Navier-Stokes approach would 
need to operate with a large set of physical parameters, 
e.g. concentrations of reacting chemical species. Thus, it 
appears that for practical values of $K$, arithmetic and 
communication demands of the Navier-Stokes models will be greater 
than of model \cite{Frankel90}. Although, complexity of 
approximation of the singular integral and the asymptotic nature 
of model \cite{Frankel90} may annul the benefit of reduction of the 
problem dimension by one.

\section{Conclusions}

Global data array transposition is the most resource consuming 
operation in simulations in question.
Developed MPI code based on a Sivashinsky type model and 
\verb#MPI_ALLTOALL# routine for transposition of column-wise 
distributed data arrays is perfectly scalable for up to a hundred 
of processors. On Cray XT4 increase of $N_{p}$ from 128 to 1024 
utilizes at least 40\% of added computational power.
Performance of the NEC SX-8 both as a massively parallel system 
and as a ``number cruncher'' is superior to other tested systems.
Achieving reasonable code efficiency requires intensive 
optimization on all considered systems.

Extrapolation of the Cray XT4 and NEC SX-8 performance shows 
that with their help reaching the stabilized acceleration regime 
of the expanding three-dimensional spherical flame numerically 
is feasible. Studies 
of the auto-transition of such flames to detonation on this 
computer are realistic as well. 
Prospects of use of massively parallel systems for more 
sophisticated asymptotic models of flame dynamics are less 
encouraging, but not hopeless.

Comparison of performances of computers was carried out on a 
particular code only and should not be interpreted as the overall 
superiority of one HPC system over another.

\subsection*{Acknowledgements}
NEC SX-8 calculations were carried out under the HPC-EUROPA project
(RII3-CT-2003-506079), with the support of the European Community - Research
Infrastructure Action under the FP6 ``Structuring the European Research Area''
Programme. 
Cray XT4 calculations were supported by the UK Consortium On Computational
Combustion For Engineering Applications (COCCFEA) under the EPSRC grant
EP/D080223.
SGI Altix 3700 resources were provided by the HPCF-UCLan at the University 
of Central Lancashire.


\begin{thebibliography}{1}
\expandafter\ifx\csname url\endcsname\relax
  \def\url#1{\texttt{#1}}\fi
\expandafter\ifx\csname urlprefix\endcsname\relax\def\urlprefix{URL }\fi

\bibitem{Gostintsev-Istratov-Shulenin88}
Y.~Gostintsev, A.~Istratov, Y.~Shulenin, A self-similar regime of free
  turbulent flame propagation in mixed gaseous mixtures., Phys. Combust.
  Explos. 24~(5) (1988) 63--70.

\bibitem{Karlin-Sivashinsky05a}
V.~Karlin, G.~Sivashinsky, The rate of expansion of spherical flames, Combust.
  Theory Model. 10~(4) (2006) 625--637.

\bibitem{Karlin-Sivashinsky05b}
V.~Karlin, G.~Sivashinsky, Asymptotic modelling of self-acceleration of
  spherical flames, Proc. Combust. Inst. 31 (2007) 1023–--1030.

\bibitem{Karlin05}
V.~Karlin, Pseudoresonant interaction between flame and upstream velocity
  fluctuations, Phys. Rev. E 73~(1) (2006) $\left.\right.$Art. No. 016305.

\bibitem{VanLoan}
C.~V. Loan, Computational frameworks for the fast fourier transform.

\bibitem{Swarztrauber98}
P.~Swarztrauber, Transposing arrays on multicomputers using de bruijn
  sequences, J. Parallel Distrib. Comput. 53~(1) (1998) 63--77.

\bibitem{AlNa'mneh-Pan_Yoo06}
R.~A. Na'mneh, W.~Pan, S.~Yoo, Efficient adaptive algorithms for transposing
  small and large matrices on symmetric multiprocessors, Informatica 17~(4)
  (2006) 535--550.

\bibitem{Frankel90}
M.~Frankel, An equation of surface dynamics modeling flame fronts as density
  discontinuities in potential flows, Physics of Fluids A~(2) (1990)
  1879--1883.

\bibitem{Greengard-Rokhlin87}
L.~Greengard, V.~Rokhlin, A fast algorithm for particle simulations, J. Comput.
  Physics 73~(2) (1987) 325--348.

\end{thebibliography}

\appendix
\section*{Appendix. The Fortran 90/MPI distributed array transposition routine}

\begin{verbatim}

MODULE sphe_d2_mpi_mach
INTEGER, PARAMETER :: ikd=KIND(1)                   ! main integers
INTEGER, PARAMETER :: rkd=SELECTED_REAL_KIND(P=15)  ! main reals
INTEGER, PARAMETER :: ikd0=KIND(1)                  ! MPI integers
END MODULE sphe_d2_mpi_mach

SUBROUTINE Transp2(u,v,nproc,Kex,nconv,nconv_mpi,ut,vt)
! Row-to-column transposition of [-Kex:nconv-Kex-1]x[-Kex:nconv-Kex-1] 
! distributed in chanks of [-Kex:nconv-Kex-1]x[1:nconv_mpi] arrays 
! over nproc processors
!
! u,v - initial and transposed arrays correspondingly
! nproc - number of processors
! Kex=2*K
! nconv=4*K+k0+1
! nconv_mpi=nconv/nproc
! ut,vt - internal work arrays
!
USE sphe_d2_mpi_mach
USE MPI
IMPLICIT NONE
INTEGER(KIND=ikd0) :: ier_mpi
INTEGER(KIND=ikd) :: j,l,nproc,Kex,nconv,nconv_mpi
REAL(KIND=rkd) :: u(-Kex:nconv-Kex-1,1:nconv_mpi)
REAL(KIND=rkd) :: v(-Kex:nconv-Kex-1,1:nconv_mpi)
REAL(KIND=rkd) :: ut(1:nconv_mpi*nconv_mpi,1:nproc)
REAL(KIND=rkd) :: vt(1:nconv_mpi*nconv_mpi,1:nproc)
!
! Executables ----------------------------------------------------
IF(nproc==1)THEN
  DO j=-Kex,nconv-Kex-1
    v(:,j+Kex+1)=u(j,:)
  END DO
 ELSE
  DO l=1,nproc
    DO j=1,nconv_mpi
      ut((j-1)*nconv_mpi+1:j*nconv_mpi,l) &
      & =u((l-1)*nconv_mpi-Kex:l*nconv_mpi-Kex-1,j)
    END DO
  END DO
  CALL MPI_ALLTOALL(ut,nconv_mpi*nconv_mpi,MPI_DOUBLE_PRECISION, &
  &                 vt,nconv_mpi*nconv_mpi,MPI_DOUBLE_PRECISION, &
  &                 MPI_COMM_WORLD,ier_mpi)
  DO j=1,nconv_mpi
    DO l=1,nproc
      v((l-1)*nconv_mpi-Kex:l*nconv_mpi-Kex-1,j) &
      & =vt(j:(nconv_mpi-1)*nconv_mpi+j:nconv_mpi,l)
    END DO
  END DO
END IF
RETURN
END SUBROUTINE Transp2
\end{verbatim}

\end{document}